\documentclass[a4paper]{article}

\usepackage[utf8]{inputenc} 
\usepackage[T1]{fontenc}    
\usepackage{hyperref}       
\usepackage{url}            
\usepackage{booktabs}       
\usepackage{amsfonts}       
\usepackage{nicefrac}       
\usepackage{microtype}      
\usepackage{xcolor}         
\usepackage{algorithmic}
\usepackage[linesnumbered,ruled,vlined]{algorithm2e}
\usepackage{bbm}
\usepackage{adjustbox}
\usepackage{booktabs}
\usepackage{wrapfig}

\usepackage{xspace}
\usepackage{graphicx}
\usepackage{capt-of}
\usepackage{varwidth}
\usepackage{subfigure}

\newcommand{\method}{\emph{Censer}\xspace}

\usepackage{INTERSPEECH2022}

\title{Censer: Curriculum Semi-supervised Learning for Speech Recognition \protect\\ Based on Self-supervised Pre-training}
\name{Bowen Zhang$^{1*}$, Songjun Cao$^{2*}$, Xiaoming Zhang$^2$, Yike Zhang$^2$, Long Ma$^2$, Takahiro Shinozaki$^1$}
\address{
  $^1$Tokyo Institute of Technology, Tokyo, Japan\\
  $^2$Tencent Cloud Xiaowei, Beijing, China}
\email{bowen.z.ab@m.titech.ac.jp}

\begin{document}

\maketitle
\def\thefootnote{*}\footnotetext{Equal contribution. This work was partly supported by JSPS KAKENHI Grand Number JP22K12069.}

\begin{abstract}
Recent studies have shown that the benefits provided by self-supervised pre-training and self-training (pseudo-labeling) are complementary. Semi-supervised fine-tuning strategies under the pre-training framework, however, remain insufficiently studied. Besides, modern semi-supervised speech recognition algorithms either treat unlabeled data indiscriminately or filter out noisy samples with a confidence threshold. The dissimilarities among different unlabeled data are often ignored.  
In this paper, we propose Censer, a semi-supervised speech recognition algorithm based on self-supervised pre-training to maximize the utilization of unlabeled data. The pre-training stage of Censer adopts wav2vec2.0 and the fine-tuning stage employs an improved semi-supervised learning algorithm from slimIPL, which leverages unlabeled data progressively according to their pseudo labels' qualities. We also incorporate a temporal pseudo label pool and an exponential moving average to control the pseudo labels' update frequency and to avoid model divergence. Experimental results on Libri-Light and LibriSpeech datasets manifest our proposed method achieves better performance compared to existing approaches while being more unified.

\end{abstract}
\section{Introduction}
\label{sec:intro}
Recent research attention in deep learning is gradually shifting towards unsupervised and Semi-Supervised Learning (SSL) where an abundant amount of unlabeled data can be utilized to improve the neural network's performance. 
Self-supervised pre-training and semi-supervised learning are two mainstreams of leveraging unlabeled data in speech recognition. Wav2vec2.0~\cite{w2v} has become the most commonly used self-supervised pre-training framework in ASR due to its competence in learning powerful audio representations. 
Semi-supervised learning approaches, on the other hand, do not require such two-stage training, but jointly train the model from scratch using both labeled and unlabeled data. A key technique in SSL is known as Pseudo-Labeling (PL, also the abbreviation of 'Pseudo Label') or Self-Training (ST), where unlabeled data are pseudo-labeled by the model itself or another teacher model. 

slimIPL~\cite{slim} is an advanced and simplified Language Model (LM) free algorithm. The core of slimIPL is that it introduces a dynamic cache to store historical information to prevent the model from over-fitting. However, either the data to add in to the cache, or the entries in the cache to use (or replace) are always randomly selected, which may cause several potential issues. First, there might be too old (never been replaced) or too new (just been replaced in the previous iteration) entries selected for training, resulting in learning low-quality PLs or over-fitting to the model's current prediction, respectively. Second, it is also hard to 
guarantee that in-cache samples and out-of-cache samples have the same overall probability to be drawn for training under such a design. 
In fact, the large amount of unlabeled data may not only contain samples that are similar to labeled data but also data points that are less homologous or with a domain shift, resulting in uneven PL qualities inferred by the model. To this end, we propose to improve slimIPL by reducing the randomness and progressively using unlabeled samples from the easiest to the hardest, similar to the idea of curriculum learning~\cite{curriculum}. 

On the other hand, recent studies~\cite{pushing, complementary} show that the benefits brought by self-supervised pre-training and ST are complementary, suggesting a way of maximizing the utilization of unlabeled data in ASR. Generally, combining the techniques involves four stages: a self-supervised pre-training stage, a fully-supervised fine-tuning stage, a PL decoding stage on the unlabeled dataset (where an LM is usually fused), and an ST stage on both labeled and pseudo labeled datasets (where parameter re-initialization is usually performed). We therefore explore unifying the last three steps with a \emph{semi-supervised fine-tuning} stage. In doing so, we show that the LM fusion and the parameter re-initialization are no longer imperatives for obtaining a strong result. 

Putting these together, we propose \method (\textbf{C}urriculum s\textbf{e}mi-supervised lear\textbf{n}ing for \textbf{s}pe\textbf{e}ch \textbf{r}ecognition), an integrated approach that maximizes the utilization of unlabeled data and simplifies the training process. The pre-training part of \method directly adopts wav2vec2.0. The semi-supervised fine-tuning part is an improved algorithm from slimIPL where the idea of curriculum learning is leveraged to progressively select unlabeled data and their PLs. 
To sum up, our contributions are two-fold:
\begin{itemize}
\item We propose a semi-supervised algorithm for ASR which progressively uses unlabeled data. It renders better performance than existing algorithms that treat unlabeled data indiscriminately or filter with a fixed confidence threshold.
\item We investigate using LM-free SSL algorithms as a semi-supervised fine-tuning stage
to replace the conventional pipeline of combining ST and pre-training. This approach gets rid of the reliance on an external LM and a re-training stage, unifying the overall training process.

\end{itemize}

\section{Related works}
Basic PL (ST) algorithms~\cite{modern,pl} in ASR use a seed model to generate PLs on unlabeled data and train a new model from scratch on the combination of labeled and pseudo-labeled data. The newly trained model can then be used as the seed model and repeat the process. Incremental PL algorithms~\cite{incremental,incremenatal2} propose to divide the unlabeled dataset into multiple splits and incrementally use these splits to constantly improve the seed model. Considering the fact that restarting from scratch for each generation is computationally heavy, iterative PL~\cite{ipl} generates PLs along with the training, simplifying the whole process. The aforementioned methods all use an LM to ensure higher qualities of PLs. This is shown in later literature~\cite{slim} to be having its disadvantages outweigh the advantages because fusing an LM does not only increase computational burden but may also lead the model to over-fit to LM knowledge. 

slimIPL~\cite{slim} is an LM-free algorithm that employs a dynamic cache to stabilize the optimization process. The cache keeps historical PLs for later use and can be updated with a certain probability. In KAIZEN~\cite{meanasr} and Momentum PL~\cite{mpl}, an Exponential Moving Average (EMA) over the model's historical weights is instead applied to avoid model divergence. There are also a number of SSL algorithms that are firstly proposed in the image recognition field~\cite{mean,mix,uda,fix,noisy} and then adopted to speech recognition~\cite{meanasr,mixasr,udaasr,fixasr,noisyasr}. Particularly, SSL algorithms with curriculum learning principles show promising results in the semi-supervised image classification field~\cite{aaai,flex}.

\cite{pushing,complementary} find that self-supervised pre-training and pseudo-labeling are complementary for speech recognition. Noisy student training~\cite{noisy, noisyasr} and a simple ST strategy are used in these works. The fine-tuned model is fused with an external LM to generate PLs for the unlabeled dataset at one stretch. The final model is obtained by re-training on the combined (labeled and pseudo-labeled) dataset.


\section{Methodology}
\subsection{Self-supervised Pre-training}
Given a labeled dataset $\mathcal{X}=\left\{x_m,y_m\right\}_{m=1}^M$ and an unlabeled dataset $\mathcal{U}=\left\{u_n\right\}_{n=1}^N$, where $N$ is usually greater or much greater than $M$, we first pre-train a model $\mathcal{M}_\theta$ only on $\mathcal{U}$ in a self-supervised fashion, and then fine-tune $\mathcal{M}_\theta$ using both $\mathcal{X}$ and $\mathcal{U}$ in a semi-supervised fashion.
For the self-supervised pre-training, we adopt wav2vec2.0~\cite{w2v}. Wav2vec2.0 pre-training allows the model to learn a strong audio representation by solving a contrastive task over quantized speech representations. The pre-training improves the ability of the seed model later used for PL, and also familiarizes the model with unlabeled data in advance for the next stage.
In the semi-supervised fine-tuning stage, the model is first trained for $S$ steps using only $\mathcal{X}$ to ensure a certain recognition ability, and then optimized on $\mathcal{X} \cup \mathcal{U}$ via SSL.

\subsection{Pseudo Label Scoring}

To reduce computational cost and avoid model over-fitting to LM knowledge as suggested in~\cite{slim}, we use only the Acoustic Model (AM) for PL generation. For our Connectionist Temporal Classification (CTC)~\cite{ctc} based AM, PLs are generated by choosing the token with the highest probability at each time step and then merging all consecutive and identical tokens. Here we consider two types of scores as the PL quality score. \\
\textbf{Confidence Score (CS)}
The merging operation in PL generation can be regarded as selecting the first tokens of each consecutive identical string and discarding the rest as the first token is decisive for the state transition; therefore, we consider the CS of a sentence as the average of posterior probabilities over these tokens. While we also tested averaging over every consecutive identical string or selecting the one with the highest probability instead of selecting the first, the performances rendered by these implementations were very close. \\
\textbf{Confidence-Robustness Score (CRS)} The prediction confidence, however, is sometimes not reliable enough as the model might be blindly confident about its predictions. Inspired by ~\cite{uncertainty}, we add a robustness score to help better assess PL qualities. Given a piece of unlabeled data $u_n$ and its PL $q_n$, we apply a weak data augmentation as a perturbation to $u_n$ and pass the perturbed version through the model to obtain $\tilde{q}_n$. We then compute the Levenshtein distance between $q_n$ and $\tilde{q}_n$ as a penalty subtracted from the confidence score. Concretely, the CRS of PL $q_n$ is computed as:
\vspace{-.05in}
\begin{equation}
\label{eq:crs}
    \mathrm{CRS}(q_n) = \frac{\mathrm{CS}(q_n) + \mathrm{CS}(\tilde{q}_n)}{2} - \lambda\frac{\mathrm{lev}(q_n, \tilde{q}_n)}{l}
\vspace{-.05in}
\end{equation}
where $\mathrm{CS}$ is the confidence score mentioned above, $\mathrm{lev}$ denotes Levenshtein distance, $l$ is the length of $q_n$ and $\lambda$ balances the weight between the two terms. The second term reflects the uncertainty of $q_n$ under perturbations; when the predictions are inconsistent, the CRS becomes low.

\subsection{Curriculum Pseudo Labeling}

\begin{figure}[t]
    \centering
    \includegraphics[width=0.47\textwidth]{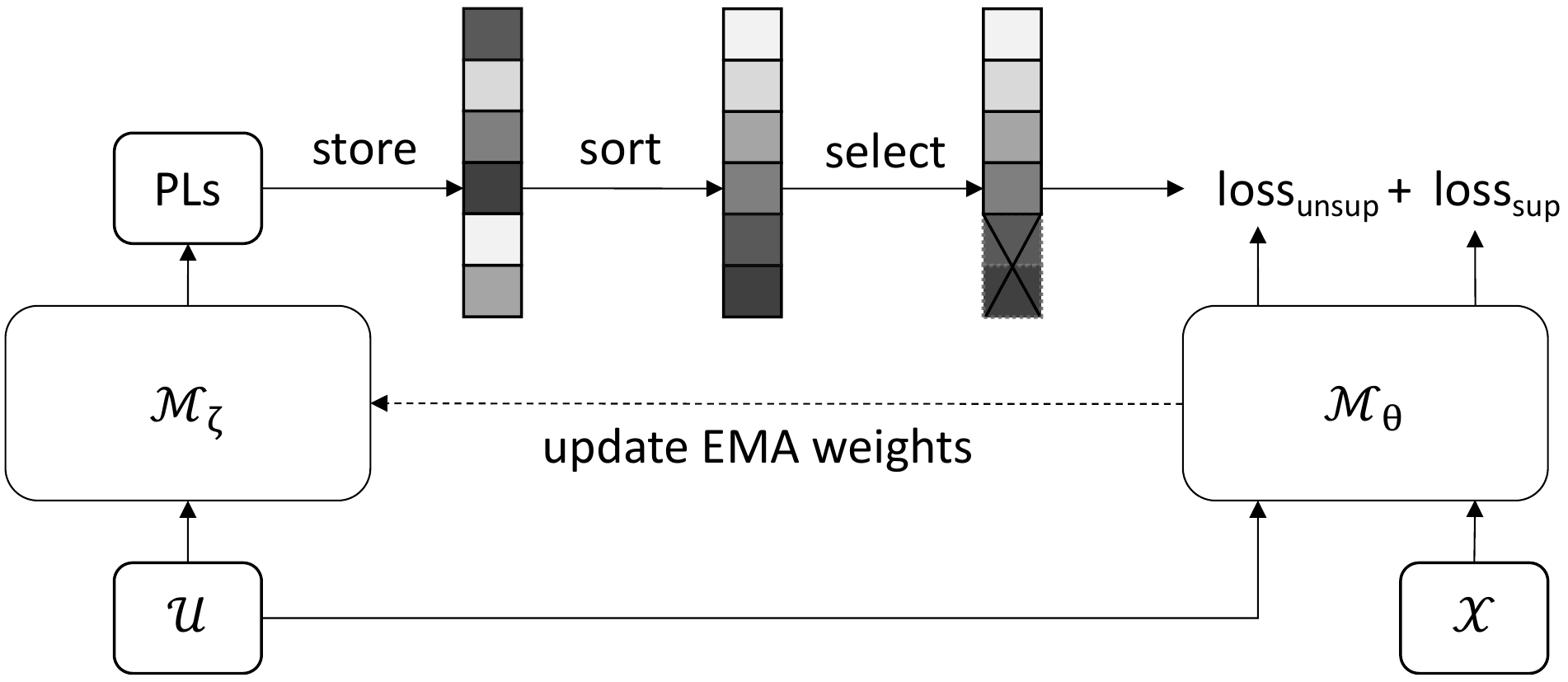}
    \caption{An illustration of the semi-supervised training process of \method. The PLs are always inferred with the EMA model, and stored and sorted in the PL pool. The best proportion of PLs are selected and used for training according to the current curriculum stage.}
    \vspace{-.2in}
    \label{fig:my_label}
\end{figure}

\textbf{Curriculum Pace} Our curriculum strategy is similar to~\cite{aaai}, where easy samples are first fitted while difficult samples are gradually added. Concretely, we divide the whole SSL training into $K$ stages; for the $k$-th stages, only unlabeled samples with top $\frac{k}{K}$ PL scores are used, while labeled data are used all along (labeled data can be considered as the easiest) with a hyper-parameter $\mu$ controlling the ratio of unlabeled samples to labeled ones in each iteration. 

However, $\frac{k}{K}$ increases as the curriculum stage $k$ proceeds, leading the number of unlabeled data to fit to increase. Setting each curriculum stage to last for the same duration in this case will make selected data of early stages to be over-fitted while those of late stages under-fitted.
To this end, the duration of the $k$-th curriculum stage is set as follows.
\vspace{-.1in}
\begin{equation}
\label{eq:k}
    \frac{k}{\sum_{k=1}^K k} \cdot F
\vspace{-.05in}
\end{equation}
Here $F$ denotes the total number of iterations in the semi-supervised training. This ensures that selected samples in different stages are iterated the same number of rounds (epochs). 
\textbf{Temporary PL Pool} In \method, instead of directly scoring the entire $\mathcal{U}$, we utilize a temporary PL pool with a capacity $C$ which is a tunable hyper-parameter, similar to the cache concept in slimIPL. This pool keeps a temporary subset of $\mathcal{U}$ and their PLs, the model will only fetch unlabeled samples from this pool. Specifically, let $\mathcal{U}_C=\left\{u_c,q_c\right\}_{c=1}^C$ be a subset sampled from $\mathcal{U}$ along with their PLs, we first sort all $\left\{u_c,q_c\right\}$ pairs in $\mathcal{U}_C$ by their PL scores in descending order to obtain $\mathcal{U}_C^{sorted}$, and keep the top $\eta$ pairs according to the current curriculum stage, resulting in $\mathcal{U}_{\eta_k}^{sorted}$, where 
\vspace{-.1in}
\begin{equation}
\vspace{-.05in}
\label{eq:eta}
    \eta_k = \frac{k}{K} \cdot C.
\end{equation}
The model will then fetch unlabeled data only from $\mathcal{U}_{\eta_k}^{sorted}$. Once all $\eta$ pairs have been used, the pool will be emptied and another $C$ samples from $\mathcal{U}$ will be drawn to the pool.

The employment of the PL pool has the following advantages:
Scores evaluated this way are more representative compared to assessing in a one-pass fashion, since the model's parameters are constantly being updated. The capacity $C$ of the pool controls an interval at which PLs are re-generated. By tuning $C$ we can control the update frequency of PLs, and a large $C$ can efficiently prevent model divergence. Also, all unlabeled data can have equal chance to be considered, since the pool is emptied when all entries in it have been used and new unlabeled data are sampled into the pool without replacement in each epoch.

\subsection{Stabilizing Model Convergence}
We found in our experiments as well as suggested in~\cite{slim,meanasr,mpl} that
the model is prone to diverge after a period of training by over-fitting to it's own predictions.
When no re-initialization is carried out, data augmentation and Exponential Moving Average (EMA) techniques become crucial.

The idea of using EMA to average a model's historical weights was first proposed in~\cite{mean}, and further explored in the ASR field in \cite{meanasr, mpl}. 
In \method, the EMA model is initialized as a copy of the AM after fine-tuning for the first $S$ steps. The EMA weights $\mathcal{M_\zeta}$ are then updated after each iteration as 
\vspace{-.05in}
\begin{equation}
\label{eq:ema}
    \mathcal{M_\zeta}=\alpha\mathcal{M_\zeta}+(1-\alpha)\mathcal{M_\theta},
\vspace{-.05in}
\end{equation}
where $\alpha$ is an EMA decay factor. During the training, PLs are always inferred with the EMA model. A large $\alpha$ reserves more historical information at inference and is of great significance in preventing model divergence. 

Finally, data augmentations are also applied to avoid over-fitting and to improve the model's robustness. There are two types of augmentations in \method: weak augmentation that uses only channel masking and strong augmentation that uses both time masking and channel masking. The masking strategy follows~\cite{w2v} which is a modified version of SpecAugment~\cite{specaug}. We apply strong augmentation to both labeled and unlabeled data during the training before feeding them to the model. The weak augmentation is used only for CRS evaluation. We also tried consistency training by letting strongly-augmented samples learn the PLs generated by their weakly-augmented counterparts as suggested in~\cite{fix,fixasr}, however, it did not bring benefits in our experiments compared to directly learning PLs generated without augmentations. Figure~\ref{fig:my_label} and Algorithm~\ref{alg:cpl} illustrate the overall process of \method.

\begin{algorithm}[tb]
\SetAlgoLined
\caption{\method algorithm.}
\KwIn{$\mathcal{X}=\left\{x_m,y_m\right\}_{m=1}^M$, $\mathcal{U}=\left\{u_n\right\}_{n=1}^N$}
\KwOut{$\mathcal{M_\theta}$}
\label{alg:cpl}
{Pre-train $\mathcal{M_\theta}$ on $\mathcal{U}$ using wav2vec2.0}\\
{Fine-tune $\mathcal{M_\theta}$ on $\mathcal{X}$ for $S$ steps}\\
{Initialize EMA weights $\mathcal{M_\zeta}=\mathcal{M_\theta}$}\\
{Draw next $C$ samples from $\mathcal{U}$}\\
{Generate PLs with $\mathcal{M_\zeta}$} and store as $\mathcal{U}_C=\left\{u_c,q_c\right\}_{c=1}^C$\\
{Sort $\mathcal{U}_C$ via (\ref{eq:crs}) in descending order}\\
{Get current curriculum stage $k$ via (\ref{eq:k})}\\
{Select the first $\eta_k$ entries from $\mathcal{U}^{sorted}_C$} via (\ref{eq:eta})\\

\While{maximum iterations not reached}{
{Randomly draw a batch $\mathcal{B}$ from $\mathcal{X}$}\\
{Fetch a batch $\mathcal{B}_U$ from $\mathcal{U}_{\eta_k}^{sorted}$} in order\\
{Apply strong augmentation to $\mathcal{B}\cup \mathcal{B}_U$}\\
{Train $\mathcal{M_\theta}$ on $\mathcal{B}^{aug}\cup \mathcal{B}^{aug}_U$}\\
{Update EMA weights via (\ref{eq:ema})}\\
\If{all samples in $\mathcal{U}_{\eta_k}^{sorted}$ have been used}
{Empty $\mathcal{U}_{\eta_k}^{sorted}$ and go back to 4}
}
\end{algorithm}

\section{Experiments}
\subsection{Experimental Setup}
\begin{table}[b]
\vspace{-.2in}
\centering
\caption{Hyper-parameters used in \method. ($S$: Number of supervised-only iterations; $F$: Number of SSL iterations measured on $\mathcal{X}$, the corresponding iterations on $\mathcal{U}$ is $\mu F$) }
\label{tb-param}
\vspace{-.05in}
\begin{adjustbox}{width=\columnwidth,center}
\begin{tabular}{lrrrrrrrr}
\toprule

Data Split & Peak lr & $S$ & $F$ & $
\mu$ & $C$ & $K$ & $\lambda$ & $\alpha$  \\  \midrule
LL-10/LS-960  & 5e-5 & 20k & 30k & 5 & 100 & 5 & 1 & 0.999960          \\

LS-100/LS-960   & 2e-5 & 80k & 80k & 1 & 100 & 5 & 1 & 0.999988     \\
\bottomrule
\end{tabular}
\end{adjustbox}
\end{table}

\begin{table*}[t]
\vspace{-.2in}
\caption{Word error rates on LL-10/LS-960 and LS-100/LS-960. Rows with $\dagger$ denote results borrowed from~\cite{mpl, slim}. LARGE denotes a larger model size as used in~\cite{slim}.}
\label{tb-main}
\vspace{-.1in}
\centering
\begin{adjustbox}{width=2\columnwidth,center}
\begin{tabular}{lrrrr|rrrr|rrrr|rrrr}
\toprule
 & \multicolumn{8}{c}{LL-10/LS-960}& \multicolumn{8}{c}{LS-100/LS-960}\\ 
 \cmidrule(lr){2-9} \cmidrule(l){10-17}
 & \multicolumn{4}{c}{AM only decoding}& \multicolumn{4}{c}{AM + LM decoding}&\multicolumn{4}{c}{AM only decoding}& \multicolumn{4}{c}{AM + LM decoding}\\
 \cmidrule(lr){2-5} \cmidrule(lr){6-9} \cmidrule(lr){10-13} \cmidrule(l){14-17}
 &\multicolumn{2}{c}{Dev. WER}& \multicolumn{2}{c}{Test WER}&\multicolumn{2}{c}{Dev. WER}& \multicolumn{2}{c}{Test WER}&\multicolumn{2}{c}{Dev. WER}& \multicolumn{2}{c}{Test WER}&\multicolumn{2}{c}{Dev. WER}& \multicolumn{2}{c}{Test WER}\\
\cmidrule(lr){2-3} \cmidrule(lr){4-5} \cmidrule(lr){6-7} \cmidrule(l){8-9}
\cmidrule(lr){10-11} \cmidrule(lr){12-13} \cmidrule(lr){14-15} \cmidrule(l){16-17}
Algorithms & \multicolumn{1}{c}{clean} & \multicolumn{1}{c}{other} & \multicolumn{1}{c}{clean} & \multicolumn{1}{c}{other} & \multicolumn{1}{c}{clean} & \multicolumn{1}{c}{other} & \multicolumn{1}{c}{clean} & \multicolumn{1}{c}{other}& \multicolumn{1}{c}{clean} & \multicolumn{1}{c}{other} & \multicolumn{1}{c}{clean} & \multicolumn{1}{c}{other} & \multicolumn{1}{c}{clean} & \multicolumn{1}{c}{other} & \multicolumn{1}{c}{clean} & \multicolumn{1}{c}{other} \\ 
\midrule\midrule

\textbf{Baselines}\\
\emph{Semi-supervised only} \\
slimIPL$^\dagger$ (LARGE)~\cite{slim}  & 11.4 & 14.0 & 11.4 & 14.7 & 6.6 & 9.6 & 6.8 & 10.5     & 3.7 & 7.3 & 3.8 & 7.5 & 2.7 & 5.5 & 3.1 & 6.2    \\ 
MPL$^\dagger$~\cite{mpl} & 8.7 & 22.0 & 9.0 & 22.4 & 6.5 & 16.9 & 6.8 & 17.1     & 8.2 & 17.5 & 8.4 & 17.6 & 6.3 & 13.5 & 6.4 & 13.7    \\ 
\midrule
\emph{Self-supervised only} \\
wav2vec2.0$^\dagger$ (LARGE)~\cite{w2v}  & 8.1 & 12.0 & 8.0 & 12.1 & 3.4 & 6.9 & 3.8 & 7.3     & 4.6 & 9.3 & 4.7 & 9.0 & 2.3 & 5.7 & 2.8 & 6.0    \\ 
wav2vec2.0~\cite{w2v}  & 9.5 & 16.5 & 9.5 & 16.7 & 4.1 & 9.6 & 4.4 & 9.9     & 6.1 & 13.6 & 6.1 & 13.3 & 2.9 & 7.9 & 3.4 & 8.0    \\ \midrule

\multicolumn{2}{c}{\emph{Combined} (w/ re-training)\;\;\;\;\;\;\;\;\;\;\;\;\;\;\;\;\;\;\;\;\;\;} \\
wav2vec2.0 + ST~\cite{complementary}  &7.9&13.5&7.9&13.3&3.6 & 8.3 & 3.9 & 8.4& 5.2&11.2&5.3&11.0&2.9&7.2&3.2&7.4  \\ \midrule\midrule

\textbf{Semi-supervised Fine-tuning}\\
\multicolumn{2}{c}{\emph{Combined} (w/o re-training)\;\;\;\;\;\;\;\;\;\;\;\;\;\;\;\;\;\;\;\;} \\
wav2vec2.0 + slimIPL   & 7.7 & 13.6 & 7.7 & 13.4 & 3.7 & 8.2 & 4.2 & 8.4 & 5.3 & 11.5 & 5.3 & 11.1 & 2.7 & 6.9 & 3.2 & 7.1    \\
wav2vec2.0 + MPL   & 7.5 & 13.2 & 7.6 & 13.0    & 3.8 & 7.8 & 4.1 & 8.1  & 5.3 & 11.4 & 5.2 & 11.2 & 2.7 & 7.0 & 3.2 & 7.2   \\ \midrule
\method (CS)   & 7.4 & 12.6 & 7.2 & 12.4 &3.8 &7.8 &4.0 & 8.1 & 5.0 & 11.0 & 5.1 & 10.8  & 2.8 & 6.8 & 3.1 & 7.1        \\
\method (CRS)   & 7.3 & 12.5 & 7.1 & 12.4  &3.8&7.8&3.9&8.1& 5.0&10.9&5.1&10.7 & 2.8 & 6.8 & 3.2 & 7.0     \\
Curriculum Oracle   & 6.9 & 12.2 & 6.7 & 12.0  &3.6&7.5&3.8&7.9& 4.9&10.7&5.0&10.5 & 2.8 & 6.7 & 3.1 & 7.0     \\
\bottomrule
\vspace{-.3in}

\end{tabular}
\end{adjustbox}
\end{table*}

\textbf{Data} All experiments are conducted using LibriSpeech (LS)~\cite{librispeech} and Libri-Light (LL)~\cite{librilight} datasets. We consider two semi-supervised setups: LL-10/LS-960 and LS-100/LS-960. These two setups use \emph{train-10h} from LL and \emph{train-clean-100} from LS as their labeled data, respectively, and both use \{\emph{train-clean-100, train-clean-360, train-other-500}\} from LS as unlabeled data. Hyper-parameters are tuned on the validation set of LS. For the self-supervised pre-training, we use the same unlabeled data (i.e. LS-960). During the whole training process, no additional resources are used (e.g. text data or an external LM). This is designed to simulate the realistic circumstances under which only a certain amount of labeled and unlabeled speech data are available, and we are encouraged to train a model by maximizing the utilization of the given resource. \\
\textbf{Model} Our model is CTC-based following the \emph{BASE} model architecture of~\cite{w2v}. The feature extractor consists of seven 512-channel 1-D convolution blocks with kernel sizes (10,3,3,3,3,2,2) and strides (5,2,2,2,2,2,2), followed by a 12-layer Transformer encoder. Each Transformer layer has self-attention dimension 768, head 8, and feed-forward dimension 3072. Finally, a prediction head with one fully-connected layer is added to map the output to vocabulary where character-level tokens are used. Raw waveform is used as the input for consistency with wav2vec2.0 pre-training. The LM used in experiments, if involved, is the off-the-shelf LibriSpeech 4-gram LM trained on a text corpus.\\
\textbf{Hyper-parameters} All models are trained using 8 Tesla V100 GPUs with a total batch size of 64. We use Adam optimizer with a tri-state learning rate schedule that gradually warms up to a peak after 10\% of total iterations, and starts to linearly decay after 50\% total iterations to 0.05 of the peak value by the end of the training. For the strong augmentation, we follow~\cite{w2v} and set time mask length to 10 time steps with a total masking probability of 0.65 and the channel mask length 64 with a probability of 0.5. For the weak augmentation, we only use channel masks with length 64 and probability 0.5.
Other hyper-parameters are listed in Table~\ref{tb-param}. Note that the pool size in Table~\ref{tb-param} is denoted in the scale of batches, when the batch size is 64, $C=100$ indicates a pool containing 6400 samples. The EMA decay factor is calculated by setting $\alpha^{F}=0.3$, meaning that by the end of the SSL training, 0.3 of the initial EMA weights are retained~\cite{mpl}.

\subsection{Experimental Results}
\textbf{Main Results}  We include semi-supervised only, self-supervised only, and the conventional combined method into our baselines. For the semi-supervised baselines, we consider two modern LM-free algorithms slimIPL~\cite{slim} and MPL~\cite{mpl}. Particularly, we also implement these two existing algorithms on top of a wav2vec2.0 pre-train model to show the universality of the semi-supervised fine-tuning approach and to keep consistent with our proposed method.
We re-implement all methods using the same model under the same codebase for a fair comparison, except for the results of a \emph{LARGE} model which are borrowed from the original papers. The combined baseline (i.e. \emph{wav2vec2.0+ST}) is re-trained for 240k iterations. The shared hyper-parameters (e.g. learning rate, etc.) are consistent among these methods, other method-specific hyper-parameters follow the original papers.

Table~\ref{tb-main} shows the main results. In the \emph{AM only decoding} column, no LMs are involved in the entire training, while the \emph{AM+LM decoding} column uses LM shallow fusion in the evaluation phase. Especially, the \emph{wav2vec2.0+ST} method uses the LM also in its third (i.e. the PL generation) stage. 
By replacing fully-supervised fine-tuning with semi-supervised fine-tuning, it brings major improvements to the \emph{wav2vec2.0} baseline, for example, the relative error rate reduction on the LL-10/LS-960 split is about 25\% when decoding with AM only. The performances of the SSL approaches are also slightly better than adding an extra ST stage to the fine-tuned wav2vec2.0 model (row \emph{wav2vec2.0+ST}), while being more unified.

Comparing among SSL methods, experimental results manifest that \method yields a stronger AM over its SSL baselines. For instance, on the LL-10/LS-960 split, the relative improvement for the AM-only experiments is about 5\%-8\% over \emph{wav2vec2.0+MPL} and \emph{wav2vec2.0+slimIPL}, thanks to the employment of the curriculum PL selection. It is shown in the next section that the curriculum PL selection is superior to filtering with a fixed confidence threshold. The improvement brought by CRS over CS, however, is not remarkable, which is because CRS is still a model's prediction confidence based scoring strategy. More sophisticated PL scoring methods can be investigated in future work. Nevertheless, to help better understand the limits of curriculum learning in semi-supervised speech recognition, we conduct a theoretical control experiment, denoted as \emph{curriculum oracle} at the bottom of Table~\ref{tb-main}. In the \emph{curriculum oracle}, the PL pool is always sorted perfectly according to the real error of the PLs, therefore, the curriculum pace proceeds perfectly from the easiest sample to the hardest. While it shows minor improvement over \method on the LS-100/LS-960 split, the improvement on the LL-10/LS-960 split is measurable, suggesting that when the number of labeled data is limited, sorting data properly and progressively using unlabeled data is of great help.\\
\textbf{Ablations} The effect of EMA-related hyper-parameters is well-studied in~\cite{mpl, meanasr}, which also applies to our experiment. Therefore, we perform ablation studies mainly on our newly introduced hyper-parameters, which are shown in Table~\ref{tb-ablation}. Specially, we add an experiment where the curriculum strategy is replaced by a confidence threshold $\tau=0.95$ to show that our progressive approach is superior to simply filtering samples with a fixed threshold.  As the results indicate, all curriculum experiments show better performance than setting a fixed threshold. As for the pool size, a small pool size containing only 10 batches results in less favorable performance; increasing the pool size to 100 reduces the error rate, but a too-large pool size does not always bring improvement because the PLs update frequency will be decreasing at the same time. Increasing the number of curriculum stages gives increasingly better performance, however, the cost of setting a large curriculum stage number is that it slows down the training in the early stages as most PLs are discarded.
\begin{table}[t]
\centering
\caption{Ablations on pool size and curriculum stage. Experiments are done on the LL-10/LS-960 split; CS is used for scoring.}
\vspace{-.05in}
\label{tb-ablation}
\begin{adjustbox}{width=\columnwidth,center}
\begin{tabular}{l|l|cccc}
\toprule

Pool size & Curriculum stage& dev-clean & dev-other & test-clean & test-other  \\  \midrule

$C=100 $&$\;\;\;\; \tau=0.95$  &7.7&13.1&7.6&13.0    \\
$C=100 $&$\;\;\;\; K=3 $ &7.5&12.7&7.3&12.4\\
$C=100 $&$ \;\;\;\;K=5 $ &7.4&12.6&7.2&12.4\\  
$C=100 $&$ \;\;\;\;K=10 $ &7.3&12.5&7.2&12.3\\ 
$C=10$&$ \;\;\;\;K=5   $ &7.5&13.0&7.4&12.7 \\
$C=1000$&$ \;\;\;\;K=5  $  &7.4&12.6&7.2&12.5  \\

\bottomrule
\end{tabular}
\end{adjustbox}
\vspace{-.25in}
\end{table}

\section{Conclusion}
This paper proposed Censer, a semi-supervised fine-tuning strategy for speech recognition. By progressively using unlabeled data from easy to hard, Censer achieves improvements over existing semi-supervised and self-training approaches. Further potentials of curriculum learning can be extracted by exploring more elaborately designed pseudo label quality scoring mechanisms.

\bibliographystyle{IEEEtran}
\bibliography{mybib}

\end{document}